\documentclass[aps,prb,reprint,superscriptaddress,floatfix]{revtex4-2}

\usepackage{dcolumn}
\usepackage{xcolor}
\usepackage{graphicx}
\usepackage{amsmath}
\usepackage{bm}
\usepackage{floatrow}
\usepackage{physics}
\usepackage{times}
\usepackage[colorlinks=true,allcolors=blue]{hyperref}%
\usepackage{relsize}
\usepackage[capitalise]{cleveref}
\DeclareUnicodeCharacter{2212}{-}

\date{\today}

\begin{document}
\title{
Time-resolved X-ray Spectroscopy from the Atomic Orbital Ground State Up
}
\author{Daniel Jost$^\ddagger$}
\email{daniel.jost@stanford.edu}
\affiliation{Stanford Institute for Materials and Energy Sciences,
SLAC National Accelerator Laboratory, 2575 Sand Hill Road, Menlo Park, CA 94025, USA}

\author{Eder G. Lomeli$^\ddagger$}
\affiliation{Stanford Institute for Materials and Energy Sciences,
SLAC National Accelerator Laboratory, 2575 Sand Hill Road, Menlo Park, CA 94025, USA}
\affiliation{Department of Materials Science and Engineering, Stanford University, Stanford, CA 94305, USA}

\author{Ta Tang}
\affiliation{Stanford Institute for Materials and Energy Sciences,
SLAC National Accelerator Laboratory, 2575 Sand Hill Road, Menlo Park, CA 94025, USA}
\affiliation{Department of Applied Physics, Stanford University, Stanford, CA 94305, USA}

\author{Joshua J. Kas}
\affiliation{Physics Department, University of Washington, Seattle, WA 98195, USA}

\author{John J. Rehr}
\affiliation{Physics Department, University of Washington, Seattle, WA 98195, USA}
\affiliation{Department of Photon Science, SLAC National Accelerator Laboratory, Menlo Park, California 94025, USA}

\author{Wei-Sheng~Lee}
\affiliation{Stanford Institute for Materials and Energy Sciences,
SLAC National Accelerator Laboratory, 2575 Sand Hill Road, Menlo Park, CA 94025, USA}

\author{Hong-Chen Jiang}
\affiliation{Stanford Institute for Materials and Energy Sciences,
SLAC National Accelerator Laboratory, 2575 Sand Hill Road, Menlo Park, CA 94025, USA}

\author{Brian Moritz}
\affiliation{Stanford Institute for Materials and Energy Sciences,
SLAC National Accelerator Laboratory, 2575 Sand Hill Road, Menlo Park, CA 94025, USA}

\author{Thomas P. Devereaux}
\email{tpd@stanford.edu}
\affiliation{Stanford Institute for Materials and Energy Sciences,
SLAC National Accelerator Laboratory, 2575 Sand Hill Road, Menlo Park, CA 94025, USA}
\affiliation{Department of Materials Science and Engineering, Stanford University, Stanford, CA 94305, USA}
\affiliation{Geballe Laboratory for Advanced Materials, Stanford University, Stanford, California 94305, USA}

\begin{abstract}
\hspace{1.7in}$^\ddagger$ \textit{These authors contributed equally.} \newline \newline
X-ray spectroscopy has been a key method to determine ground and excited state properties of quantum materials with atomic specificity. Now, new x-ray facilities are opening the door to the study of pump-probe x-ray spectroscopy - specifically time-resolved x-ray absorption (trXAS) and time-resolved resonant inelastic x-ray scattering (trRIXS). In this paper we will present simulations of each of these spectroscopies using a time-domain full atomic multiplet, charge transfer Hamiltonian, adapted to study the properties of a generalized cluster model including a central transition metal ion caged by ligand atoms in a planar geometry. The numerically evaluated trXAS and trRIXS cross-sections for representative electron configurations $3d^9$ and $3d^8$ demonstrate the insights that can be obtained from charge transfer pumping, and how this nonequilibrium process affects ground and excited state properties. The straightforward characterization of the excitations in these systems, based on our analysis of the simulations, can serve as a benchmark for future experiments, as access to these time-resolved spectroscopic techniques becomes more widely available.
\end{abstract}

\maketitle

\section{Introduction}
\label{sec:intro}

X-ray spectroscopies at resonant absorption edges have a long-standing reputation as a cornerstone in the atom specific interrogation of ground and excited state properties of strongly correlated oxides. Using well-know resonances and different polarization geometries, x-rays have revealed the coordination numbers of atoms in solids \cite{EXAFS}, the nature of spin or orbital ground state configurations \cite{XMCD}, as well as numerous excitations reflecting multiplet, charge transfer, orbital, magnetic, and lattice collective motion as a function of momentum transfer over large portions of the Brillouin zone \cite{RIXS, Dean:2013, LeTacon:2014, Jia:2016, Chaix:2017}. With increased resolving power and the development of modern, high-throughput synchrotron beamlines, x-ray spectroscopies such as x-ray absorption spectroscopy (XAS) and resonant inelastic x-ray scattering (RIXS) have matured into well-utilized techniques applicable to the study of many correlated and functional materials~\cite{LeTacon:2011, Ghiringhelli:2012, Hu:2014, Miao:2018, Hepting:2018, House:2020, Kang:2020, Hepting:2020, Lu:2021, Gilmore:2021, Ueda:2023, Schueler:2023}. 

A new frontier in time-resolved x-ray spectroscopy has slowly been built off of the foundation of time-resolved optical studies of materials~\cite{Dienst:2013, Stojchevska:2014, Kubacka:2014, Gerber:2015, Gerber:2017, Dean:2016, Li:2019, Mitrano:2020, Wang:2020, Wandel:2022, Baykusheva:2023}. The development of free electron lasers, especially those with high repetition rate capabilities needed for x-ray spectroscopies, has begun to open new windows into the behavior of correlated materials. For example, studies of the excited state properties of small molecules and atomic clusters, utilizing time-domain atomic force microscopy \cite{AFM}, now have been extended into the x-ray regime using femtosecond XAS~\cite{trFDG,Durr,Mitrano,Higley:2019} and x-ray photoemission spectroscopy (XPS)~\cite{puntel:2023} to reveal detailed information about the dynamics of excited states in correlated materials used for photocatalysis or water splitting~\cite{Bergmann:2021}. Recent time-resolved RIXS studies in iridates have shown the first ever view of magnetic short-range correlations within a photo-excited ultrafast transient state~\cite{Dean}. In the near term, trXAS and trRIXS are poised to become broadly used, exciting new tools with application in diverse fields of materials science, touching on the identification and design of new photocatalysts, optimizing component battery materials for improved electrochemical cycling and capacity, and in general understanding collective dynamics of materials driven or operating out of equilibrium~\cite{Tsutsui:2021,Tsutsui:2023}.

In this work, we will focus on an application of these advanced spectroscopies for the study of strongly correlated electron systems~\cite{Basov:2011, Fradkin:2012}. The smallest building blocks of many of these materials are made up of $3d$ transition metals, surrounded by O ligand atoms. These systems often feature ground states that lie energetically close to each other~\cite{Corboz:2014}, making them sensitive to (external) perturbations. In contrast to the standard methods to tune these different ground states, such as chemical substitution, optical pumps can be used as elegant non-invasive and versatile perturbations. We will show that the pump polarization is a critical tuning parameter enabling dynamic symmetry breaking and how that is reflected in the X-ray probe response of trXAS and trRIXS. Furthermore, we will show how these probes allow us to map out excited state properties inaccessible in equilibrium settings. 
 
\section{Theory of Time-Resolved Spectroscopies}

The theory for time-resolved spectroscopies involves a straightforward extension of the familiar equilibrium tools, but cast in a time-domain formalism that does not require time-translation invariance. Specifically, for trXAS with an incident x-ray having momentum ${\bf q}$, energy $\omega$, and polarization $\epsilon$, the expression can be written as~\cite{Yuan}
\begin{eqnarray}
    \mu_{{\bf q},\omega}(t) &=& 
    \int_{-\infty}^{\infty}  dt_1 \int^{\infty}_{-\infty} dt_2 s^t_{\bf q}(t_1)s^t_{\bf q}(t_2) e^{i\omega(t_2-t_1)}\nonumber\\
&&\times \langle \mathcal{\hat D}^\dagger_{{\bf q,\epsilon}}(t_2) \mathcal{\hat D}_{\bf q,\epsilon}(t_1) \rangle,
    \label{eq:XAS}
\end{eqnarray}
where the dipole transition operator $\hat D$, selected by the x-ray polarization ${\bf \epsilon}$, is defined as 
\begin{align*}
\mathcal{\hat D}_{\bf q,\epsilon}(t)=\sum_{\mu,\nu,{\bf k},\sigma}
D_{\mu,\nu,\sigma}^\epsilon({\bf q},t) \hat c_{{\bf k+q},\mu,\sigma}^\dagger(t) \hat d_{\bf k,\nu,\sigma}(t).
\end{align*}
Here, $\hat d_{{\bf k}, \mu, \sigma}(t)$ removes a core electron at time $t$ having orbital index $\mu$, momentum ${\bf k}$ and spin $\sigma$, and $\hat c^\dagger $ likewise creates a valence electron with the specified quantum numbers. The shape function $s^t_{\bf q}(t')$ reflects the temporal envelope, or profile, of the probing x-rays, independent of the pump field. In what follows, we take the shape function to be a normalized Gaussian independent of momentum as $s_{\bf q}^t(t')=s^t(t')=\textrm{exp}[-(t-t')^2/\tau^2]/(\sqrt{2\pi}\tau)$, having a width $\tau$, which sets the overall probe time resolution.

As a resonant scattering technique, trRIXS contains unequal-time photon-in and photon-out processes for incident and scattered photons having momenta ${\mathbf{q}_{i,s}}$, energies $\omega_{i,s}$, and polarizations $\epsilon_{i,s}$, respectively. The cross-section is given by a time-domain Kramers-Heisenberg expression~\cite{Yuan}
\begin{eqnarray}\label{eq:RIXScrossSec}
&&\mathcal{I}({\mathbf{q}_i, \mathbf{q}_s,\omega_i,\omega_s;\epsilon_i,\epsilon_s})(t)\!\nonumber\\
&=&\int_{-\infty}^\infty dt_2 \int_{t_2}^{\infty} dt_2^\prime \int_{-\infty}^{\infty} dt_1 \int^{\infty}_{t_1}\!dt_1^\prime 
e^{i\omega_i(t_2-t_1)-i\omega_s(t_2^\prime-t_1^\prime)} 
\nonumber\\
&&  s_{{\mathbf{q}_i}}(t_1)s_{\mathbf{q}_i}(t_2)s_{\mathbf{ q}_s}(t_1^\prime)s_{\mathbf{q}_s}(t_2^\prime) \times\nonumber\\
&&\big\langle \hat{\mathcal{D}}_{\mathbf{q}_i,\epsilon_i}^\dagger(t_2)\hat{\mathcal{D}}_{\mathbf{q}_s,\epsilon_s}(t_2^\prime) \hat{\mathcal{D}}_{\mathbf{q}_s,\epsilon_s}^\dagger(t_1^\prime)   \hat{\mathcal{D}}_{\mathbf{q}_i,\epsilon_i}(t_1)  \big\rangle.
\end{eqnarray}
The four-time RIXS correlation function describes the propagation of the full many-body electronic state including the core and valence levels.

An applied pump field can be incorporated as a time-dependent vector potential coupling to a gauge degree of freedom of the electrons in the solid. To date, experimentally these pump fields typically are in the optical range of frequencies, modulating charge transfer in the valence band of electrons. The time-dependence of the pump field enters via the Heisenberg representation of the time-evolution of all operators appearing in Eqs.~(\ref{eq:XAS}) and (\ref{eq:RIXScrossSec}).

\section{Method}
\label{sec:method}

\begin{figure}
    \centering
    \includegraphics[width=86mm]{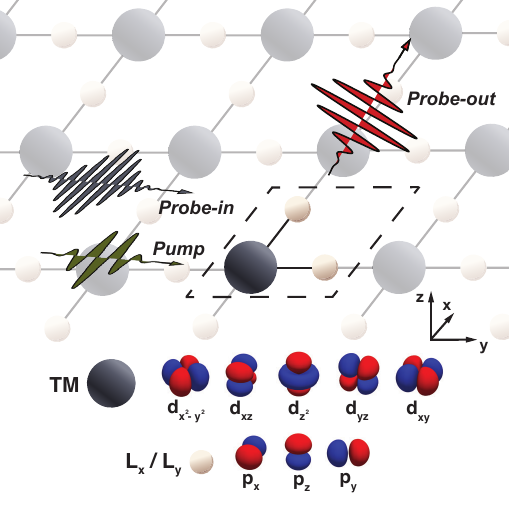}
    \caption{Sketch of the cluster model (dashed box) used for the numerical simulations according to Eqn. \ref{eq:hamiltonian}, with the transition metal (TM) $3d-$orbitals bonded with the ligand $2p-$orbitals $L_x$ along the $x-$direction and $L_y$ along $y-$direction, forming a corner sharing bonding pattern. The field direction $\hat{\varepsilon}$ of the pump is always set to point along the TM-$L_x$ bonds.}
    \label{Fig0:cluster}
\end{figure}

Here, Eqs.~(\ref{eq:XAS}) and (\ref{eq:RIXScrossSec}) are evaluated using time-evolved exact diagonalization of clusters formed from TMO$_2$ unit cells (with transition metals TM), each consisting of 5 $3d$-orbitals and 2 ligand O, each with 3 $2p$-orbitals, oriented $90^\circ$ to one another with respect to the transition metal atom, giving a total of 11 orbitals per unit cell (\textit{cf.} Figure~\ref{Fig0:cluster}). We will investigate clusters having the two distinct orbital configurations $3d^9$ and $3d^8$, i.e. with one and two holes in the unit cell, respectively. The equilibrium Hamiltonian is
\begin{align}
    \hat H &= \sum_{i,j,\sigma}\sum_{\mu,\nu} t_{i,j}^{\mu,\nu} \hat c^\dagger_{i,\mu,\sigma}
\hat c_{j,\nu,\sigma} \nonumber\\
& + \frac{1}{2}\sum_{i,\sigma,\sigma'} \sum_{\mu,\nu,\mu',\nu'} U_{\mu,\nu,\mu',\nu'} \hat c^\dagger_{i,\mu,\sigma}\hat c^\dagger_{i,\nu,\sigma'}\hat c_{i,\mu',\sigma'}\hat c_{i,\nu',\sigma} \nonumber\\
& + \frac{1}{2}\sum_{i,\sigma,\sigma'} \sum_{\mu,\nu,\mu',\nu'} U_{\mu,\nu,\mu',\nu'} \hat c^\dagger_{i,\mu,\sigma}\hat d^\dagger_{i,\nu,\sigma'}\hat c_{i,\mu',\sigma'}\hat d_{i,\nu',\sigma} \nonumber\\
& - \sum_{i,\sigma,\sigma'} \sum_{\mu,\nu} \lambda_{\mu,\nu}^{\sigma,\sigma'} \hat d^\dagger_{i,\mu,\sigma} \hat d_{i,\nu,\sigma'},
\label{eq:hamiltonian}
\end{align}
with $t_{i,j}^{\mu,\nu}$ the orbital-dependent valence electron hopping, $U$ the full on-site Coulomb matrix elements, including interactions on both the transition metal and ligand sites, as well as core-valence interactions, and $\lambda$ the core electron spin-orbit coupling. In what follows, typical parameters for charge-transfer insulators \cite{parameters} have been chosen, with only the on-site orbital energies set differently for the two cases.

A driving pump field ${\bf A}({\bf r},t)$ is included via a Peierls substitution \cite{Peierls:1933} whereby a time-dependence enters via $t_{i,j}\rightarrow t_{i,j}e^{i\phi_{i,j}(t)}$ with the Aharonov-Bohm phase \cite{Aharnonov:1959} given as $\phi_{i,j}(t)=-e \int_{\bf j}^{\bf i} d{\bf r} \cdot {\bf A}({\bf r},t)$ (units $\hbar=c=1$). We note that this specific form of coupling has limitations given that the phase does not depend upon basis orbitals used and does not represent certain types of transitions made possible via $\mathbf{A}(\mathbf{r},t)$, such as dipole-allowed transitions between orbitals located on the same site. This is a complicated affair related to gauge-invariance which has been further investigated in Ref.~\cite{gauge}. However, gauge-invariance can be satisfied, since the Hamiltonian does not contain dipole-allowed transitions on the same site. The dipole operator matrix elements include the pump field
\begin{eqnarray}\label{eq:dipoleMatrixElement}
    D_{\mu,\nu}^\epsilon({\bf q},t)     &= &-\frac{e}{2m}\sqrt{\frac{1}{2\epsilon_0\omega_{\bf q} V}} \int d{\bf r}~ \psi^*_{\mu}({\bf r}) e^{i{\bf q\cdot r}} \nonumber\\
    &&\times\epsilon\cdot \left[i\nabla + \mathbf{A}({\bf r},t)\right] \psi_{\nu}({\bf r}),
\end{eqnarray}
with $\psi_{\nu}$ the basis orbitals, $V$ the unit cell volume, and $\omega_{\bf q}$ the photon energy. Given that our consideration will be x-ray spectroscopy driven by an optical pulse, modifications of the x-ray dipole transition by the pump field have been neglected in the these matrix elements. 

For simplicity, the vector potential for the pump field is chosen as ${\bf A}({\bf r},t)=\hat \varepsilon A_0\sin(\omega_{\textrm{pump}}t)e^{-t^2/2\Gamma_p^2}$, where $A_0$ sets the pump amplitude, and the spatial dependence is negligible, since the optical wavelength is much larger than the lattice scale. Throughout, we work with the pump frequency $\omega_{\textrm{pump}} = \pi\,\mathrm{eV}$ and pulse width $\Gamma_p^2=1.5\,\mathrm{eV}^{-2}$. The pump frequency is chosen to reflect the fact that a larger reorientation of charge is obtained for smaller pump field frequencies \cite{Ta}. Finally, the field direction $\hat\varepsilon$ is set to point along the $x$-TM-ligand bonds, breaking the $C_4$ symmetry of the cluster while the pump is on.

\section{Results}
\label{sec:results}

\subsection{Orbital occupations}

\begin{figure}[h!]
\centering
\includegraphics{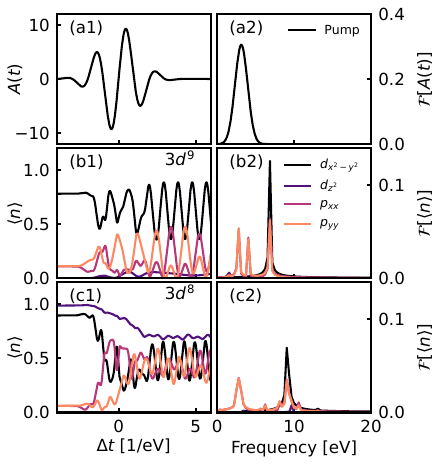}
    \caption{\label{Fig1} \textbf{Time Dependent Ground State Occupations} -- (a1) Optical pump field with frequency $\omega_{\textrm{pump}}  = \pi\,\mathrm{eV}$ centered at $\Delta t = 0$ \cite{parameters} and (a2) its Fourier transform $\mathcal{F}[A(t)]$.  
    (b1) $3d^9-$cluster orbital $\langle n \rangle$ occupation as a function of $\Delta t$ and (b2) its Fourier transform, where $d_{i}, i \in [{x^2-y^2}, z^2]$, denote the TM $3d-$orbitals and $p_{\alpha \alpha}, \alpha \in [x, y]$, denote the ligand $2p-$orbitals. (c1) Time-dependent $3d^8-$cluster orbital occupation and (c2) its Fourier transform. 
    }
\end{figure}

To establish a baseline for results from the RIXS calculations, we first examine the effect of the pump field on the orbital occupation of each cluster. Figure~\ref{Fig1}(a1) presents the profile of the pump field centered at time delay $\Delta t = 0$. Panel (a2) depicts the Fourier transform $\mathcal{F}$ showing a singular peak at $\pi$\,eV with which we establish the unit-correspondence between the time and frequency domains. The resulting orbital occupations $\langle n \rangle$ for the $3d^9-$ and $3d^8-$clusters are shown in Figure~\ref{Fig1}(b1)-(c2), {\it i.e.} for one and two holes in the unit cell, respectively.

For times $\Delta t \ll 0$, the ground state configurations of the two example clusters are predominantly of spin-$1/2$ $3d^9$ and spin-$1$ $3d^8$ character, with admixtures of holes in the $\sigma-$bonded oxygen orbitals, denoted as $2p_{\alpha\alpha}$, with $\alpha$ indicating both the orbital and direction with respect to the transition metal in the unit cell. Other hole weights are negligible and not plotted here. The pump field, having polarization along the $x-$direction causes charge transfer between the transition metal and the bonding oxygen orbitals. The amount of charge transfer following the application of the pump is controlled by the magnitude as well as the frequency of the applied field, as noted previously~\cite{Ta}. 

Evident from the hole redistribution in the $3d^9-$cluster at the rising edge of the pump field for $\Delta t \rightarrow 0$, the hole weight reduction in the TM $3d_{x^2-y^2}$ orbital bleeds mainly into the oxygen $2p_{\alpha\alpha}$ orbitals, with a smaller transfer towards the $3d_{z^2}$ orbital. After this initial redistribution of the hole weight and in the absence of the pump field, {\it i.e.} for times $\Delta t \gg 0$, the cluster settles into a non-equilibrium state with oscillations in the orbital occupations $\langle n \rangle$. The average occupation of the $3d_{x^2-y^2}$ decreases and a singular frequency of about 6.8\,eV dominates the long-time behavior of the TM $3d_{x^2-y^2}$ occupation [Figure~\ref{Fig1}(b2)], associated with the charge transfer onto the ligand orbitals. As expected, the $2p_{\alpha\alpha}$ oscillation shows one component at this same frequency, and two additional frequency components indicate the presence of ligand-to-ligand charge transfer for which the orbital occupations oscillate out-of-phase [panel (b1)].

A qualitatively similar picture emerges in the case of the $3d^8-$cluster, in which the hole weights for $\Delta t \ll 0$ are almost evenly distributed between the $3d_{x^2-y^2}$ and $3d_{z^2}$ orbitals owing to the splitting between the $e_g-$ and $t_{2g}-$orbitals, with small weight on the ligands. Upon pumping ($\Delta t \rightarrow 0$), the hole weight on both $3d_{e_g}$-orbitals drops, with a weak oscillation setting in for $3d_{z^2}$, and the hole weight is redistributed to the $2p_{\alpha\alpha}$ orbitals. The long-time behavior, evident from $\mathcal{F}[\langle n \rangle]$, is dominated essentially by two frequencies: one associated with the charge transfer setting the interchange of hole weight between $3d_{x^2-y^2}$ and $2p_{\alpha\alpha}$ orbitals and a second indicating charge transfer between the oxygen ligand orbitals, with a weak temporal modulation on the $3d_{z^2}$ orbital.  

The results summarized in Figure~\ref{Fig1} deliver the following insights: as the pump forces hole weight to redistribute from the TM to the $\sigma-$ bonded ligand orbitals, it creates a pump-induced excited state
whose state vector changes periodically in time, yielding the continuous oscillations of the hole occupation. These oscillations can be viewed as a manifestation of multi-level Rabi oscillations for which the pump-pulse plays the role of quantum state preparation. 

\subsection{Time-resolved x-ray spectroscopy}
\label{subsec:trXRS}

The oscillating occupation of the TM and ligand orbitals should be reflected in the trXAS and trRIXS intensities for both TM clusters. To track the field-induced transfer of charge between the transition metal and the ligand, trXAS and trRIXS are examined by evaluating Eq.~\ref{eq:XAS} and \ref{eq:RIXScrossSec}, respectively, for the parameters used to obtain Fig.~\ref{Fig1}.

\subsubsection{$3d^9$ cluster}
\label{subsubsec:3d9_results} 

\begin{figure}
    \centering
    \includegraphics{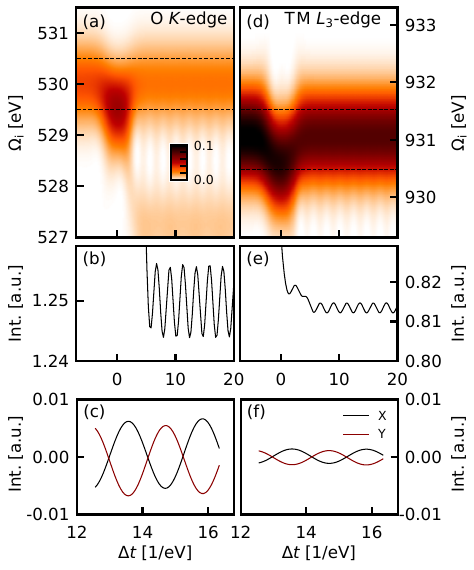}
    \caption{\textbf{3\textit{d}$^\mathbf{9}$ trXAS for O \textit{K}-edge (left panels) and TM \textit{L}$_\mathbf{3}$-edge (right panels)} -- Spectra as a function of $\Delta t$.
     (a)~O $K-$edge results, showing a softening of the XAS peak at $\Delta t \sim 0\,\mathrm{eV}^{-1}$, which recovers for $\Delta t \gg 0$, and the emergence of a weak side-band that develops 2.5 eV below the main peak. (b) Integrating the intensity in the vicinity of the equilibrium main peak at the O $K-$edge (between the black dashed lines in (a)) reveals oscillations for times $\Delta t \gg 0$, which are out-of-phase for incident X and Y polarization, as shown in panel (c). Analogous results are obtained at the TM $L_3-$edge: (d) peak energy softening at $\Delta t \sim 0\,\mathrm{eV}^{-1}$, (e) weak oscillations of the main peak, and (f) a dichroic signal for X and Y polarization. 
     }
    \label{Fig2:XAS_CuO}
\end{figure}

Figure~\ref{Fig2:XAS_CuO} depicts the simulated trXAS of the $3d^9-$cluster at the O $K-$edge [panels (a) to (c)] and TM $L_3-$edge [panels (d) to (f)], with X-ray polarization along the $x-$axis. While the pump field is on, the trXAS intensity at the O $K-$edge softens, reflecting the inversion of the hole occupation from the formerly occupied metal and ligand states. The trXAS signal partially recovers at long times, recovering the feature centered at around 530 \,eV in the equilibrium XAS. A peak emerges at lower incident energies, sitting approximately 2.5\,eV below the main peak. This pre-edge feature indicates the formation of pump-induced Zhang-Rice singlet (ZRS) states~\cite{Zhang_Rice:1988}. 

The multi-level Rabi oscillations and concomitant hole weight modulations lead to corresponding fluctuations of the trXAS intensity at both edges, observable 
at the longer oscillatory periods of the O charge density, as seen in Figure~\ref{Fig1}. To highlight these oscillations, we show the integrated intensity around the main peak at the O $K-$edge in Fig.~\ref{Fig2:XAS_CuO}(b). These oscillations, and the overall decrease of the main peak intensity, follow from a shifting hybridization between the metal and ligand. Similar to the O $K-$edge data, the TM $L_3-$edge signal softens and recovers [Fig,~\ref{Fig2:XAS_CuO}(d)], with intensity fluctuation at long times [Fig.~\ref{Fig2:XAS_CuO}(e)], albeit having a weaker amplitude than at the O $K-$edge. The drop of intensity in the main peak at both edges following the pump pulse can be attributed to the formation of ZRS states, which opens an additional channel for the absorption process.  

A polarization analysis of the trXAS intensity, again integrated over the main peak, provides more insight and facilitates a direct comparison of the long-time behavior revealed by the shifting orbital occupations in Fig.~\ref{Fig1}. Focusing on a time window where the initial effects of the pump pulse have subsided, the oscillations, as shown in Figs.~\ref{Fig2:XAS_CuO}(c) and (f), appear to be out-of-phase between X and Y polarized X-ray photons. The persistence of the dichroic signal, even at long times after the pump, shows that the system retains memory of the broken $C_4-$symmetry induced during the duration of the pump pulse, due to its well-defined polarization. This long-lived effect underscores the lasting influence of the pump on the electronic structure of the cluster. 

\begin{figure}
    \centering
    \includegraphics{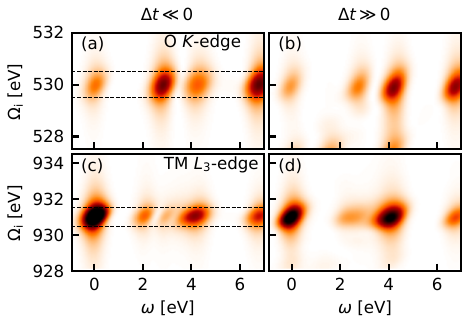}
    \caption{\textbf{3\textit{d}$^\mathbf{9}$ trRIXS before and after the pump} -- Incident energy maps for $\Delta t \ll 0$, $\Delta t \gg 0$ across the O $K-$ and TM $L_3-$edge in panels (a),(b) and (c),(d), respectively. The dashed lines in each panel indicate the energy range of the main peaks at the O $K-$edge or TM $L_3-$edge, respectively (as in Fig.~\ref{Fig2:XAS_CuO}). Polarization for all maps lies along the x--axis for both incoming and outgoing photons.}
    \label{Fig3:3d9_fullmaps}
\end{figure}

Additional details about the pump-induced changes of the electronic properties can be revealed by trRIXS. Unlike XAS, which provides information about the core-excited (intermediate) state, RIXS extends this by offering insights into the final state properties. This distinction allows us to disentangle the nuanced effects of optical pumping on the excited state spectrum (i.e. electronic properties reflected in charge-transfer and orbital $dd-$excitations), elements that are convoluted within the full XAS response. 

Figure~\ref{Fig3:3d9_fullmaps} shows a qualitative comparison of trRIXS, with full incident energy $\Omega_\mathrm{i}$ versus energy loss $\omega$ maps, across the O $K-$edge and TM $L_3-$edge for two times: $\Delta t \ll 0$, before the pump pulse, and $\Delta t \gg 0$, well after the pump pulse. At the O $K-$edge, the intensity which bleeds into the lower side-band of the trXAS signal stems, at least partially, from a redistribution of spectral weight associated with a high energy excited state at an energy transfer $\sim4$\,eV. This excitation is associated with charge transfer of the $\pi_z-$bonding orbital. Further contribution to this side-band originates from a fainter signal at an energy transfer of $\sim1.8$\,eV, which is not visible in equilibrium ($\Delta t \ll 0$) for this polarization channel. At the TM $L_3-$edge the spectral weight redistribution is limited to the incident energy range corresponding to the XAS main peak. 

\begin{figure}
    \centering
    \includegraphics{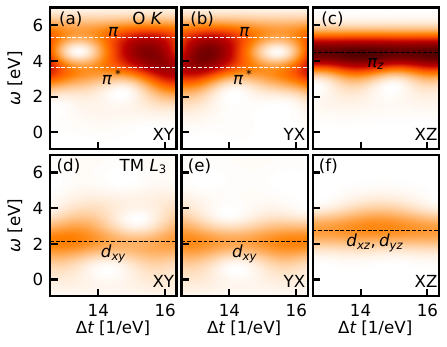}
    \caption{\label{Fig4:3d9_KL} 
    \textbf{3\textit{d}$^\mathbf{9}$ trRIXS for O \textit{K}-edge (top panels) and TM \textit{L}$_\mathbf{3}$-edge (lower panels) at different crossed polarization} --
    Integrated trRIXS intensities for the incident energy windows shown in Figs.~\ref{Fig2:XAS_CuO} and \ref{Fig3:3d9_fullmaps}, across the main XAS peaks. (a)-(c) O $K-$edge energy loss versus time maps with dashed lines corresponding to energies of the $\pi^*-$ and $\pi-$ charge transfer excitations, showing out-of-phase intensity distributions for XY and YX polarizations. (d)-(f) TM~$L_3-$edge energy loss versus time maps for the indicated polarization channels. Intensity modulation of the orbital excitations is less pronounced than for ligand-to-ligand charge transfer. 
    }
\end{figure}

Figure~\ref{Fig4:3d9_KL} shows the trRIXS intensity, as a function of energy loss versus time, integrated within windows around the XAS main peaks at the O $K-$ and TM $L_3-$edges (the dashed lines in both  Figs.~\ref{Fig2:XAS_CuO} and \ref{Fig3:3d9_fullmaps}) for parallel and crossed polarization channels. 
The full excitation spectrum accessible via the second dipole transition in the photon scattering process helps to identify the individual components predominantly responsible for the oscillations in the trXAS signal. The ability to tune the incoming and outgoing polarization provides a means of discriminating between charge-transfer and different orbital or $dd-$excitations. 
For example, in Figs.~\ref{Fig4:3d9_KL}(a) and (b), integrated trRIXS at the O $K-$edge reveals charge transfer to the $\pi^*-$anti-bonding and $\pi-$bonding ligand orbitals. These excitations are not seen at the TM $L_3-$edge, instead contrasted with the visible $dd-$excitations having $d_{xy}-$character. Generally, the modulation of intensity in the orbital excitations at the TM $L_3-$edge is less pronounced than the ligand-to-ligand charge-transfer excitations at the O $K-$edge, mirroring the decreased intensity in the trXAS signal shown in Figure~\ref{Fig2:XAS_CuO}. The observed intensity modulation between the orbital excitations at the TM $L_3-$edge and the ligand-to-ligand charge transfer excitations at the O $K-$edge highlight that the primary action occurs on the O atoms in the $3d^9-$cluster and suggest the pump-induced formation of a ZRS state due to photo-doping, consistent with side-band trace seen in the trXAS~\cite{Chen:2013}. This indicates that optical pumping can serve as a reliable and non-invasive method to photo-dope systems, presenting an alternative to traditional chemical doping methods, which often introduce disorder. The ability to dynamically induce such states with optical pumping opens prospects for experiments aimed at driving dynamic phase transitions and exploring the intricate dynamics of charge transfer between ligands in strongly correlated systems.  

\subsubsection{$3d^8-$cluster}
\label{subsubsec:3d8}
\begin{figure}
    \centering
        \includegraphics{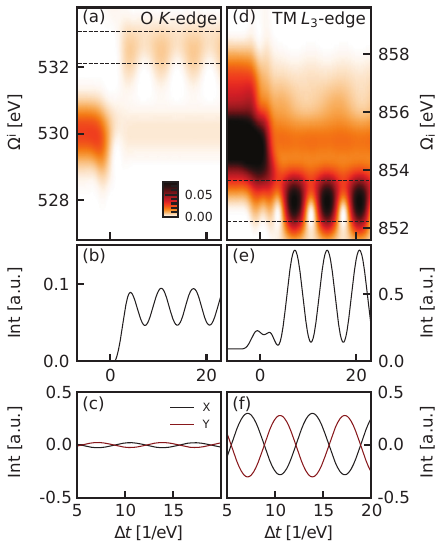}
    \caption{\label{Fig5:3d8KXAS} 
    \textbf{3\textit{d}$^\mathbf{8}$ trXAS for O \textit{K}-edge (left panels) and TM \textit{L}$_\mathbf{3}$-edge (right panels)} --
    (a)~O $K-$ edge XAS as a function of time delay $\Delta t$. 
    Weak intensity appears at $\sim 3\,\mathrm{eV}$ above the main peak. (b) Integrated intensity within the dashed lines in panel (a), showing oscillations in the side-band intensity. (c)~Side-band O $K-$edge intensities exhibiting out-of-phase oscillations for X and Y polarization. (d)~TM $L_3-$edge XAS as a function of $\Delta t$, showing a drop in the intensity of the main peak at $\sim 855\,\mathrm{eV}$ following the pump pulse, which is redirected to lower incident photon energies at $\sim 853\,\mathrm{eV}$. (e)~Integrated intensities within the dashed lines in (d) showing the oscillations of the pump-induced side-band. (f)~Side-band TM $L_3-$edge intensities oscillating out-of-phase for X and Y polarization. 
    }
\end{figure}

TM-complexes hosting a $3d^8$ configuration can serve as quintessential examples of charge-transfer insulators~\cite{Zaanen:1985}, such as NiO, or building blocks of nickelate superconductors~\cite{Li:2019,Li:2020}, pivotal for elucidating the electronic structure in strongly correlated electron systems~\cite{Shen:1990, Shen:1991}. With nominally two holes in the $3d-$manifold and split $e_g-$ and $t_{2g}-$orbitals~\cite{parameters}, Hund's coupling facilitates a high-spin $S=1$ state, making the $3d^8-$cluster  an intriguing laboratory to investigate the balance between inter- and intra-atomic exchange energy scales. The simulated trXAS signals at the O $K-$ and TM $L_3-$edges are depicted in Fig.~\ref{Fig5:3d8KXAS}. In the center of the pump pulse at $\Delta t = 0$, the spectral weight of the main O $K-$edge XAS peak located at 530\,eV drops significantly [panel~(a)]. That weight recovers only partially at longer times $\Delta t \gg 0$, highlighting the long-term pump-induced effects on the occupation of the $2p-$orbitals. The pump opens up a side-band, sitting roughly 3\, eV above the main absorption peak exhibiting slow intensity oscillations. Integrating over a window encapsulating this side-band in panel~(b) shows these weak oscillations. The main peak at the TM $L_3-$edge has a drop in intensity at $\Delta t = 0$, which remains weak for long-times [$\Delta t \gg 0$, as shown in panel (c)]. The side-band seen here develops about 2\,eV below the XAS main absorption peak, and integrating over a window across the side-band yields a time-trace exhibiting strong oscillatory behavior [panel~(d)], comparable in magnitude to the main XAS peak in equilibrium.  

Similar to the $3d^9-$cluster, a polarization analysis of the trXAS shows the time-dependent dichroism of the upper side-band at the O $K-$edge and the lower side-band at the TM $L_3-$edge [Figs.~\ref{Fig5:3d8KXAS}(c) and (f)]. These out-of-phase oscillations indicate the symmetry breaking character of the optical pump surviving in the long-time limit: these results show how polarization-dependent trXAS can be used as an effective probe to investigate whether a polarized pump induces dynamical symmetry breaking. This phenomenon underscores the potential of polarization-dependent trXAS as a sensitive probe to discern the effects of a polarized pump, particularly in distinguishing trivial phenomena like heating from more complex dynamics such as photo-induced phase transitions. Further, contrasting these oscillations with those of the $3d^9-$cluster, it seems reasonable that stronger covalency with increased charge delocalization leads to slower oscillations, which makes the observation of these oscillations experimentally feasible using state-of-the-art trXAS.   

\begin{figure}
    \centering
    \includegraphics{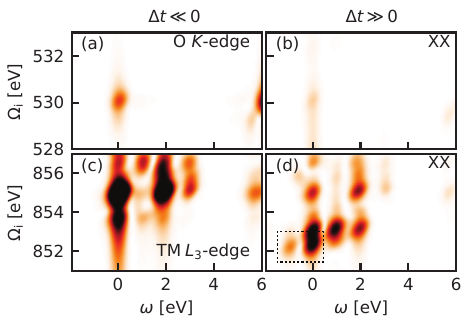}
    \caption{\label{Fig6:3d8_fullmaps} 
    \textbf{3\textit{d}$^\mathbf{8}$ trRIXS before and after the pump} --
    (a),(b) Incident photon energy versus energy loss maps at the the O $K-$edge for two times $\Delta t \ll 0$ and $\Delta t \gg 0$. Following the pump ($\Delta t \gg 0$), the incident photon energy dependence of trRIXS smears out across the O $K-$edge, lacking any well-defined features in this energy range. (c),(d) Across the TM $L_3-$edge, trRIXS as a function of incident photon energy versus energy loss 
    shows excitations from the elastic line up to 2 eV energy transfer. A photo-induced anti-stokes (energy gain) feature can be seen at $\sim -0.9\,\mathrm{eV}$ energy transfer.}
\end{figure}

To elucidate the nature of the sideband 2\,eV below the main XAS peak at the TM $L_3-$edge, we turn to the trRIXS data sown in Figure~\ref{Fig6:3d8_fullmaps}. Prior to the pump pulse, the TM $L_3-$edge intensity sits predominantly at $\sim 855$\,eV with strong $dd-$excitations, discernible at an energy loss of 2\,eV. Post-pump, there is a conspicuous redistribution of spectral weight, extending 2 eV below the equilibrium resonance. Concurrently, an additional feature emerges at 1\,eV, corresponding to the site energy difference between the $e_g-$ and $t_{2g}-$orbitals. On the energy gain side, the appearance of an anti-Stokes feature at 0.9\,eV is a sign that the pump disrupts the equilibrium electronic state. 

\begin{figure}
    \centering
    \includegraphics{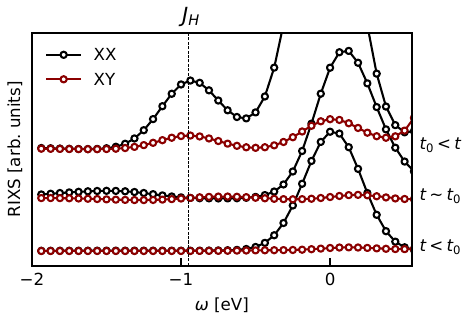}
    \caption{\label{Fig7:AS}
    \textbf{3\textit{d}$^\mathbf{8}$ anti-Stokes feature in trRIXS} --
    Polarization dependence of the integrated intensity within the dashed box in Figure~\ref{Fig6:3d8_fullmaps} for three selected times. Before the pump pulse at $t<t_0$ the response on the energy-gain, {\it i.e.} anti-Stokes, side is flat. At the center of the pump pulse, at $t\sim t_0$, the intensity increases only marginally. At long times ($t_0 < t$), a feature, stronger in XX polarization than in XY polarization, becomes visible and has a maximum at $\sim 0.9\,eV$.}
\end{figure}

The anti-Stokes feature exhibits a polarization dependence, as illustrated in Fig.~\ref{Fig7:AS}, which shows cuts of the side-band below the TM $L_3-$edge at three different representative times before, during, and after the pump. Before the pump, the anti-Stokes (energy gain) side shows a uniform response, as expected for an equilibrium spectrum. During the pump, there is a slight increase in intensity, signalling the onset of pump-induced changes in the electronic structure. However, these changes are not pronounced, suggesting that the immediate effect of the pump on the energy dynamics of the $3d^8-$cluster is relatively modest. 

For long times, with the pump off, a noticeable feature develops, predominantly in XX polarization. This feature, which is more distinct than in XY polarization, has a maximum at 0.9\,eV on the energy gain side, corresponding roughly to the energy scale of the Hund's exchange energy $J_H$ and the 10D$_q$ crystal field splitting of 1.0 eV used in the calculation. This anti-stokes transition arises from a non-equilibrium ground state consisting partly of a superposition of $3d^8$ configurations including either a double occupancy in one of the $e_g$ orbitals or an additional hole in the $t_{2g}$ orbitals. These contributions, accessible via the dynamical pump-excited state, manifest themselves in RIXS as an anti-stokes $dd-$ excitation. Those involving a singlet to triplet transition can be selectively accessed via cross-polarized measurements, as opposed to non-spin transforming $dd-$contributions dominating in parallel polarization. The potential diversity in new excitations accessed in similar $3d^8$ NiO-type materials merits further investigation, where probing this dynamic electronic ground state will manifest features beyond the spin constrained stokes transitions at equilibrium. The capability of these probes to map the eigenspectrum and to deduce Hamiltonian specific parameters, as well as characteristic quasiparticle dynamics in these systems, affords time-resolved spectroscopies a unique window of opportunity to impact fundamental materials studies.

\section{Discussion and Outlook}
We compare our results to recent experimental observations: Baykusheva \textit{et al.}~\cite{Baykusheva:2023} reported a softening of the Cu $L_3-$edge trXAS response in a hole-doped cuprate while the O $K-$edge signal remained by and large unchanged. The redshifted Cu $L_3-$edge is qualitatively similar to our findings in the pure $3d^9-$cluster, close to $\Delta t \sim 0$. While this agreement is encouraging and suggests that our approach can connect directly to experimental data, we note that a quantitative comparison of O $K-$edge is not yet possible at this stage, as the formalism introduced here does not yet include a treatment to account for doping in cuprate materials which affects the features seens at the O $K-$edge more dramatically. 

Regarding the relatively fast timescales on which we observe changes in the trXAS response, a reflectivity study by Dal Conte \textit{et al.}~\cite{DalConte:2015} reports on a fast buildup of the scattering rate due to electron-boson coupling on a timescale of about 15 fs. This timescale is comparable to our findings and insinuates that a similarly fast response could be inherent to correlated materials. 

We also note that an extension of our approach to include multiple sites may enable a direct comparison with studies such as Wang \textit{et al.}~\cite{Wang:2022} who explored the ultrafast manipulation of antiferromagnetic order in NiO via sub-gap optical excitation.

Given the current state-of-the-art, the framework we present here provides a robust starting point for investigating a wide range of geometrical and electronic configurations in transition metal oxides. Our approach is designed to be versatile and adaptable, and can be extended to a wide range of systems: it can serve as a starting point for superconducting cuprates and nickelates, where one can initially constrain the parameter space from the equilibrium XAS and RIXS response, in conjunction with, by now, widely accessible experimental data. Using our formalism, it is then straightforward to obtain a quantitative assessment of the upper bound of the timescales which determine the dynamic response.  

\section{Conclusion}

Time-domain simulations of XAS and RIXS for two transition metal oxide complexes, with one and two holes in the unit cell serving as generalized $3d^9$ and $3d^8$ systems, respectively, have provided insights on the ways in which nonequilibrium processes impact ground and excited state properties. 
When driven out of equilibrium, Rabi oscillations develop in these spectra, attributed to a superposition of multi-particle eigenstates with time-dependent coefficients. Both trXAS and trRIXS can be used to selectively project the wave function of these time-evolved states, and yield information about the dynamics of ligand-to-metal and ligand-to-ligand charge transfer. Our results highlight the complex nature of charge transport in highly correlated electron systems, and demonstrate that pumping can be used to dynamically break rotational symmetry, of which the electronic system retains memory for extended periods of time.

Accessible excitations in and out of equilibrium are distinct. In equilibrium, the system resides exclusively in the ground state. In contrast, the pump-excited state can probe anti-stokes excitations, thereby allowing for direct access to important physical properties and materials parameters. In the case of the $3d^8-$cluster, trRIXS reveals intriguing insights by highlighting the energy scale of the Hund's exchange $J_H$, providing a potential way to probe a material's Hundness~\cite{Yin:2011}. 

Our study showcases the capability of trRIXS to access nonequilibrium quantum states and serves as a powerful lens through which the intricate properties of quantum materials can be examined: as we demonstrate, trRIXS serves as a direct and selective probe for the coefficients of time-dependent many-body wave functions, reflecting the underlying quantum properties in a unique manner. By offering insights into how these coefficients evolve over time, trRIXS enables a precise understanding of nonequilibrium states of matter. This approach opens avenues for comprehensive exploration of the equilibrium and dynamical properties of quantum materials.  

\section*{Acknowledgments}
This work was supported by the U.S. Department of Energy (DOE), Office of Basic Energy Sciences, Division of Materials Sciences and Engineering. Computational work was performed on the Sherlock cluster at Stanford University and on resources of the National Energy Research Scientific Computing Center (NERSC), using NERSC award BES-ERCAP0027203, supported by the U.S. DOE, Office of Science, under Contract no. DE-AC02-05CH11231. 

\bibliography{TRCTHFAM}

\end{document}